\documentclass[a4paper,11pt]{article}
\usepackage{pos}
\usepackage{graphicx}
\usepackage{tabularx}
\usepackage{booktabs}
\usepackage{multirow}
\usepackage{hepunits}
\usepackage{slashed}
\usepackage{rotating}
\usepackage{tikz}
\usetikzlibrary{decorations.pathreplacing,calc,tikzmark,patterns}
\newcommand{\MeV}{\,\mathrm{MeV}}

\newcommand{\fm}{\,\mathrm{fm}}

\title{Charmed semileptonics with twisted-mass valence quarks}

\author*[a]{Julien Frison}
\author[b]{Gregorio Herdo\'iza}
\author[b]{Carlos Pena}
\author[b]{Jos\'e \'Angel Romero}
\author[b]{Javier Ugarrio}

\affiliation[a]{ZPPT/NIC, DESY Zeuthen,\\
  Platanenallee 6, 15738 Zeuthen, Germany}

\affiliation[b]{Instituto de Fisica Te\'orica UAM-CSIC,\\
  C/ Nicolás Cabrera 13-15, Campus de Cantoblanco, 28049 Madrid, Spain}

\emailAdd{julien.frison@desy.de}

\abstract{
  Our charm program uses a mixed action with twisted-mass valence quarks over non-perturbatively improved Wilson sea quarks, in order to study various quantities in a relativistic and manifestly local framework of full QCD. The sea sector consists of $N_\mathrm{f}=2+1$ ensembles generated by the CLS initiative. Taking advantage of open boundary conditions, this allows access to fine ensembles without topological freezing. Here we focus in particular on our current progress on $D\to K\nu l$ and $D\to \pi\nu l$ semileptonics. Those are first and foremost useful for the computation of the CKM matrix elements $|V_{cs}|$ and $|V_{cd}|$.
We show that all discretisation effects seem to be reasonably under control with this choice of action, in particular those related to hypercubic lattice artefacts. Eventually, we obtain preliminary results of the form factors as a very smooth curve on the whole range of momentum transfer, and in particular the signal at zero $q^2$ appears to have the potential to be competitive with earlier published results.
}

\FullConference{%
 The 38th International Symposium on Lattice Field Theory, LATTICE2021
  26th-30th July, 2021
  Zoom/Gather@Massachusetts Institute of Technology
}

\begin{document}
\maketitle


\section{Introduction}

Flavour physics, and in particular heavy flavour physics, is nowadays one of the most active fields of activity in which physics beyond the Standard Model is expected to be observable in the near future.
Additionally, most of the fundamental parameters of the Standard Model are flavour-related, and extracting them with the highest possible precision is always desirable. Several experiments are making progress
in this direction, such as LHCb and Belle II for the bottom sector or BESIII for the charm sector. They are matched by similar progresses in the theory, where Lattice Field Theory is taking more and more importance. However, many challenges are still ahead and this field is known for some longstanding 'puzzles' such as inclusive-exclusive tensions.

In this work we are first and foremost trying to extract the CKM matrix elements $|V_{cd(s)}|$ from $D\to\pi(K)\nu l$ semileptonic decays. Semileptonics are one of the main constraints on these matrix elements, together with the related leptonic decays studied in another subproject \cite{Conigli:2021}. We aim at covering directly the full range of decay kinematics, without relying on model-dependant extrapolations, and doing
so with a relativistic and manifestly local lattice action, which offers a cross-check to the dominating staggered results.

At a previous conference \cite{Frison:2019doh} we presented our framework and sketched a strategy. It involves using a twisted mass term in the valence action, both for the light and heavy quarks, while our
CLS $N_{\rm f}=2+1$ configurations contain non-perturbatively improved Wilson light fermions. This gives us some remnants of automatic $O(a)$ improvement, guaranteeing the absence of $O(am_c)$ discretisation
terms, while preserving the non-perturbative improvement in the sea sector, protecting us from isospin breaking effects in the sea, using only renormalisation factors we already computed, and last but not least not having to generate dedicated configurations. The parameters of the light sector are matched as explained in \cite{Herdoiza:2021}.

\section{Contraction strategy}

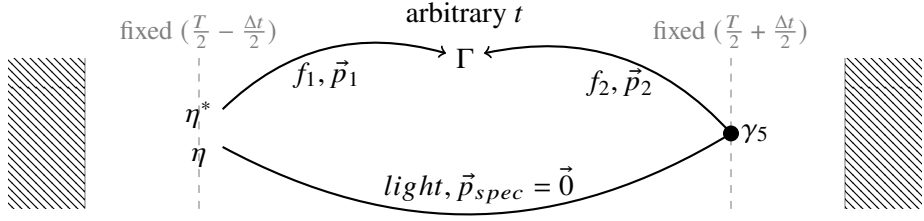
\begin{figure}{H}
\begin{center}
\begin{tikzpicture}
  \node (source) at (-3.5,0.0){$\begin{matrix}\eta^*\\\eta\end{matrix}$} ;
  \draw[-,thin,dashed,gray] (-3.5,-1.0) -- (-3.5,1.0) node[above] {\small fixed $(\frac{T}{2}-\frac{\Delta t}{2})$};
  \node (sink) at (3.5,0.0){} ;
  \draw[-,thin,dashed,gray] (3.5,-1.0) -- (3.5,1.0) node[above] {\small fixed $(\frac{T}{2}+\frac{\Delta t}{2})$};
  \fill [black] (sink) circle (3pt)
	    node[right]{$\gamma_5$};
  \node (operator) at (0.0,1.0){$\Gamma$} ;
  \draw (operator.north) node[above] {arbitrary $t$} ;
  \draw[-,gray] (-5.0,-1.0) -- (-5.0,1.0);
  \fill[gray,pattern=north west lines] (-5.0,-1.0) rectangle (-6.0,1.0);
  \draw[-,gray] (5.0,-1.0) -- (5.0,1.0);
  \fill[gray,pattern=north west lines] (5.0,-1.0) rectangle (6.0,1.0);
  \path[->,thick] (source) edge[bend left=30] node[pos=0.5,below] {$f_1,\vec p_1$} (operator);
  \path[<-,thick] (operator) edge[bend left=30] node[pos=0.5,below] {$f_2,\vec p_2$} (sink.center);
  \path[-,thick] (sink.center) edge[bend left=30] node[pos=0.5,above] {$light,\vec p_{spec}=\vec 0$} (source);
\end{tikzpicture}
\end{center}
  \caption{\label{fig:contraction}Contraction for the 3-point function. $f_1$ and $f_2$ run over any choice of flavour while $\vec{p_1}$ and $\vec{p_2}$ take a few values between $0$ and $\pm 700\ \MeV$. $\Delta_t$ is typically fixed to $2\ \fm$, which offers a good compromise between excited state contamination and signal precision, while staying far enough from the open boundaries in time. $\eta$ represents a $Z_2$ noise while $\gamma_5$ multiplies the spectator propagator to form a source for the sequential. Each inversion provides values for any $t$ and $\Gamma$.}
\end{figure}

Extracting matrix elements typically requires 3-point functions, with asymptotically large time separation between those points. This must be performed for a reasonable cost and with a sufficiently good signal-to-noise ratio, since an exact computation of all 3-point functions with naive all-to-all propagators would be way beyond the capabilities of current supercomputers. We therefore select a limited subset of time separations between the two mesons, a limited subset of momentum choices injected through twisted boundary conditions, and then use three well-established techniques which are stochastic, sequential propagators and the one-hand-trick. Additionally, distance preconditioning (DP) is used to improve the convergence of heavy quark inversions and avoid floating-point errors, so that we keep a good signal at moderately large times. This results in contractions such as Fig.~\ref{fig:contraction}. The inversions are performed with deflation in the light sector, so that the light and heavy propagators have a similar cost.

\section{Ensembles and parameters}

We use $N_{\rm f}=2+1$ CLS ensembles along the $Tr[M]={\rm cst}$ line of physics and with open-boundary conditions. This choice of boundary is important to have access to very fine ensembles where the
topological tunneling can quickly become an issue, and results in the lattice unnornmalised charm mass being as low as $0.14$, as presented in Tab.~\ref{tab:ensembles}.
This technique allows for even finer ensembles such as J500 ($0.039\ \fm$) which is now available as well and is likely to be added to this project.

In this proceedings we choose to set the focus on a subset of our ensembles which will have a particularly strong influence on the final results. It goes along two lines which can in large part be considered independently:
\begin{itemize}
  \item First, we need to make sure the discretisation effects are under control, which is easier to look at along the $m_{ud}=m_s$ line. This is the most important challenge and a test of our framework.
  \item Then, we also want to see how strong the $M_\pi$ dependence is, which is easier to look at on the coarsest ensembles and without changing any other parameter.
\end{itemize}
Given the current statistics and our computing resources, the other ensembles are unlikely to provide more than bounds on other subleading effects of lesser interest.

\begin{table}[tb]
\begin{center}
\begin{tabular}{ccccccccccc}
\toprule
  id &   $a$[fm] &  $N_\mathrm{s}$  &  $N_\mathrm{t}$  & $m_\pi$[MeV] &   $m_K$[MeV] &  $m_\pi L$ & $\Delta t$[fm] & $a\mu_c$ \\
 \midrule
  H101 & 0.086 & 32 & 96	& 420 &420  & 5.8 & 1.3, 1.5, 2.0 & 0.22 \\
  H102 & 0.086 & 32 & 96	& 350 &440  & 4.9 & 1.5 & 0.22 \\
  H105 & 0.086 & 32 & 96	& 280 &460  & 3.9 & 1.5, 2.0 & 0.22 \\
\midrule
  H400 & 0.076 & 32 & 96	& 420	&420  & 5.2 & 1.9, 2.7 & 0.21 \\
\midrule
  H200 & 0.064 & 32 & 96	& 420	&420  & 4.4 & 2.0, 2.6 & 0.18 \\
\midrule
  N300 & 0.050 & 48 & 128	& 420	&420  & 5.1 & 1.9, 2.5 & 0.14 \\
\bottomrule
\end{tabular}
  \caption{\label{tab:ensembles}Ensembles for which we present preliminary results in this proceedings. A few other ensembles have been analysed but are not reported here. In most cases only a single $Z_2$ noise
  has been used.}
\end{center}
\end{table}

\section{Parametrisation}

The matrix elements between states of the channels of interest are extracted from the 3-point functions
\begin{equation}
  C_{D\to\pi,\Gamma}(t,\Delta_t) = \sum_{m,n} \langle 0\mid \bar u\gamma_5 d\mid\pi,m\rangle  \langle\pi,m\mid\bar d\Gamma c\mid D,n\rangle \langle D,n\mid\bar c\gamma_5 d\mid 0\rangle e^{-E_{\pi m}(\Delta t-t)-E_{D n}t}
\end{equation}
and we will simply write (in the {\it physical basis} of twisted mass fermions)
\begin{equation}
  \langle S \rangle = \langle\pi,m\mid \bar dc\mid D,n\rangle \qquad\textrm{and}\qquad \langle V_\mu\rangle = \langle\pi,m\mid\bar d\gamma_\mu c\mid D,n\rangle .
\end{equation}

We use two methods to extract those matrix elements. The main one consist in totally ignoring all excited states, after having chosen $\Delta t$ sufficiently large, and then building the double ratios
\begin{eqnarray*}
  \left|\langle S \rangle\right|^2 &=& 4E_DE_\pi \frac{C_{D\to\pi,1}(t,\Delta_t)C_{\pi\to D,1}(t,\Delta_t)}{C_{D\to D,1}(t,\Delta_t)C_{\pi\to\pi,1}(t,\Delta_t)} \\
  \left|\langle \hat V_\mu \rangle\right|^2 &=& 4p_Dp_\pi \frac{C_{D\to\pi,\gamma_\mu}(t,\Delta_t)C_{\pi\to D,\gamma_\mu}(t,\Delta_t)}{C_{D\to D,\gamma_\mu}(t,\Delta_t)C_{\pi\to\pi,\gamma_\mu}(t,\Delta_t)} ,
\end{eqnarray*}
where the hat stands for the renormalised quantity, which is automatically obtained thanks to charge conservation in the denominator.

This ratio is symmetric by construction, and the values on the plateau are so strongly auto-correlated that one can simply take the value at the middle time, rather than performing a fit.

On the ensemble H101 we also generated data for several values of $\Delta t$ and applied a combined multiexponential fit directly on the 2-point and 3-point functions.
The correlated $\chi^2$ is computed for many models and plateaus and fed to a Bayesian average \cite{Jay:2020jkz}.
The comparison will be discussed very shortly in Sec.~\ref{sec:time}.

Another test is offered by the computation of the Ward identities relating $\langle S\rangle$ and $\langle\hat V_\mu\rangle$.

Once we have extracted the matrix elements, the last step is to solve for the physical form factor given by Lorentz symmetry:
\begin{eqnarray*}
  \langle S\rangle &=& \frac{M_D^2-M_\pi^2}{\mu_c-\mu_l} f_0(q^2)\\
    \langle\hat V_\mu\rangle &=& \left[P_\mu - q_\mu \frac{M_D^2-M_\pi^2}{q^2}\right] f_+(q^2) + q_\mu \frac{M_D^2-M_\pi^2}{q^2} f_0(q^2)
\end{eqnarray*}

\begin{figure}[ht]
\begin{center}
\begin{tikzpicture}
    \draw[->] (-1,0) -- (5,0) node[right] {$\Re q^2$};
    \draw[->] (0,-1) -- (0,1) node[above] {$\Im q^2$};
    \draw[<->,thin] (0,3pt) -- (2,3pt);
    \draw (1,3pt) node[above] {physical $q^2$};
    \draw (2,1pt) -- (2,-3pt) node[anchor=north] {$t_-$};
    \draw (3,1pt) -- (3,-3pt) node[anchor=north] {$t_+$};
    \draw[ultra thick] (3,0) -- (4.8,0);
    \fill [gray] (3.2,0) circle (4pt)
	    node[above]{$M_{D^*}$};
\end{tikzpicture}
\end{center}
  \caption{\label{fig:zexp}Sketch of the analytical structure on which the $z$-expansion is based.}
\end{figure}
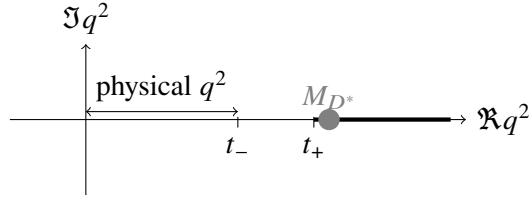

Additionally, one can perform a $z$-expansion of those form factors, which is a conformal mapping of the cut of Fig.~\ref{fig:zexp} into a circle centered around some $t_0$:
\begin{equation*}
  z(q^2,t_0) = \frac{\sqrt{t_+-q^2}-\sqrt{t_+-t_0}}{\sqrt{t_+-q^2}+\sqrt{t_+-t_0}}
\end{equation*}
While we only need very tiny interpolations between the many kinematics for which we have data, performing them in the $z$ space is cleaner and makes any model-dependent assumption completely negligible.

\section{Preliminary results}

\subsection{Results on a coarse ensemble}

\begin{figure}[ht]
  \begin{center}
    \includegraphics[width=0.8\textwidth]{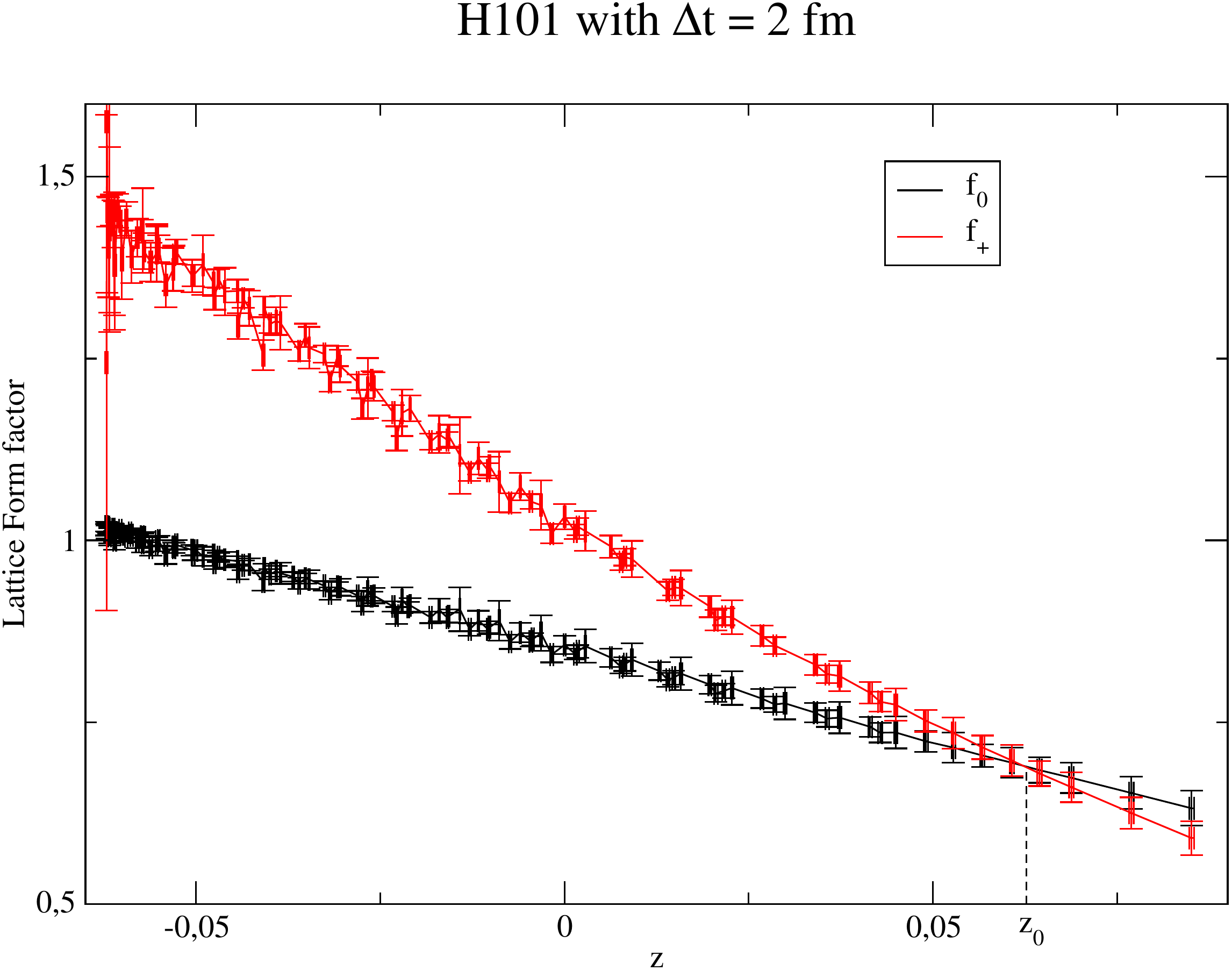}
  \end{center}
  \caption{\label{fig:H101deltaT}We represent all the points for which we have a direct computation of form factors for H101, which are a product of the choices of momenta on $D$ and $\pi$. The large number of kinematics
  make it obvious that our curves are very smooth and any Lorentz-breaking term (which could arise at $O(a^2)$ or higher) is absent. The value $z=z_0$ where the curves cross correspond to $q^2=0$. Despite injecting large momenta to reach this kinematics, the signal did not degrade much.}
\end{figure}

The coarsest ensembles are not only the cheapest but also a great benchmark for the discretisation effects of our mixed action: if either $O((am_c)^n)$ or $O((ap)^n)$ terms get out of control, this would
first be visible on our coarsest ensembles, where this effect is the largest while the noise is the smallest. 
The form factors are shown in Fig.~\ref{fig:H101deltaT} using the double ratio and the largest $\Delta_t$.

\subsection{On momentum dependence}
\begin{figure}[ht]
  \begin{center}
    \includegraphics[width=0.5\textwidth]{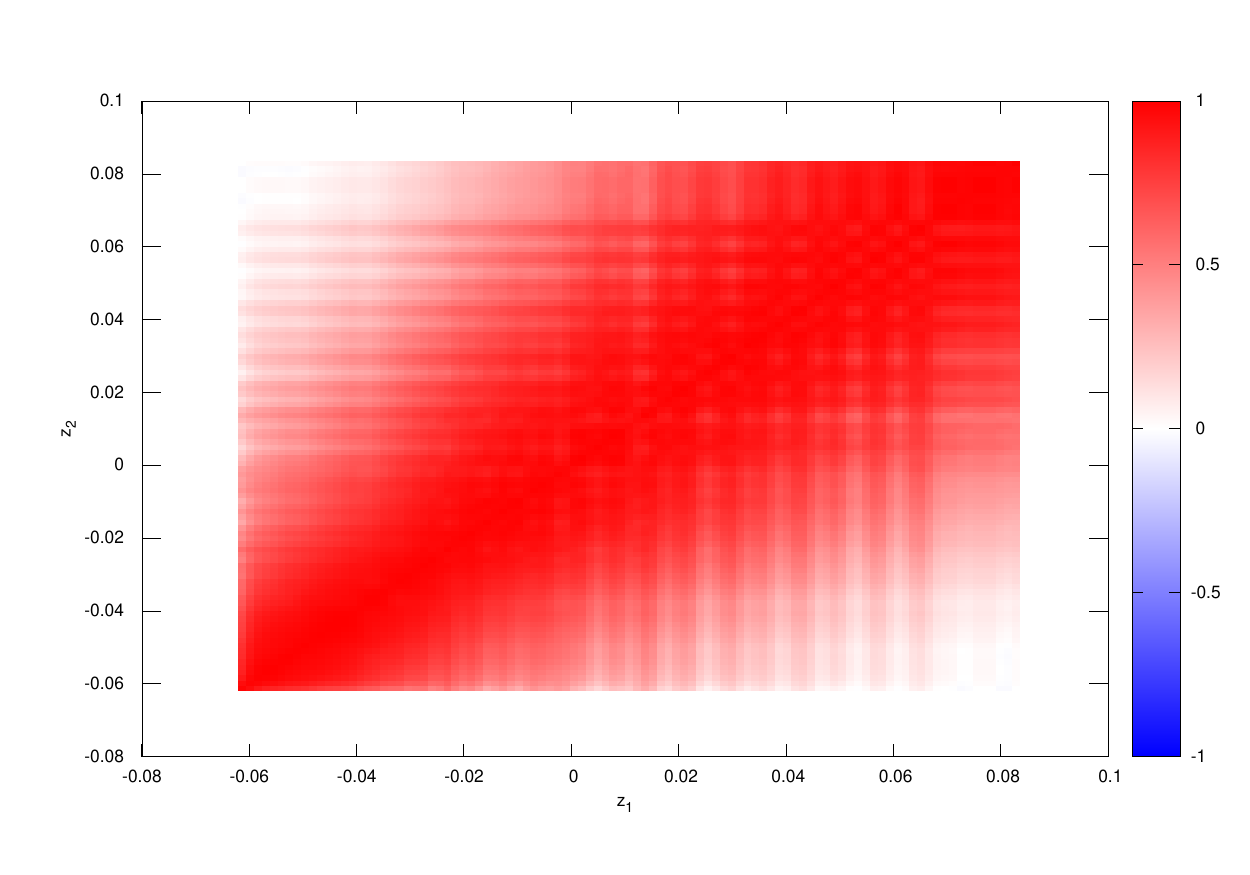}
  \end{center}
  \caption{\label{fig:momcorr}Correlation of the form factor $f_+$ (available experimentally) between two values of $z$ called $z_1$ and $z_2$. This is obtained after some smoothing and re-sampling of the interpolated form factor, at a scale much smaller than the correlation length.}
\end{figure}

Our use of many momenta was driven by a few considerations:
First we did not know whether reaching directly $q^2=0$ would be possible, with good statistics and a good control of discretisation, nor if hypercubic effects \cite{Lubicz:2016wwx} would need to be dealt with, and
we were optimistic about how much can be gained by our contraction strategy which gets $O(N^2)$ correlators for the cost of $O(N)$ inversions.
Describing the curve directly on the such a dense set also means we are sure we are not introducing any model-dependent error, and allows to exploit all of the available experimental data.
It eventually turned out that those points are strongly correlated, as shown in Fig.~\ref{fig:momcorr}.

\subsection{The choice of time separation}
\label{sec:time}
An inconvenient of our contraction strategy
is that the time separation between the two mesons needs to be chosen in advance and cannot be varied without extra inversions.
While distance preconditioning and working in a pseudoscalar meson channel reduce the impact of having to choose a large $\Delta t$ to eliminate excited states,
it still has some impact on the precision we can reach.
Based first on 2-point runs and then on our first 3-point runs, we discovered that choosing this separation around $2\ \fm$ seems to give some reasonable compromise and fit on all lattices. While it is
not yet fully clear whether a single-state analysis has a negligible bias with such a $\Delta t$ or whether a multi-exponential analysis will be needed, this bias is for sure not much greater than our target precision.
However, if we reduce $\Delta t$ below this value, at $1.5$ or even $1.3\ \fm$, we get a clear signal of something going wrong: the curves of the form factors are not smooth anymore, the time-dependent
contamination of excited states breaks Lorentz symmetry. This is particularly visible on $\langle V_i\rangle$ and $f_+$ at largish $q^2$, and shows striking tensions even when the double ratio still appears to
produce plateaus with a good $\chi^2$.

\begin{figure}[ht]
  \begin{center}
    \includegraphics[width=0.6\textwidth]{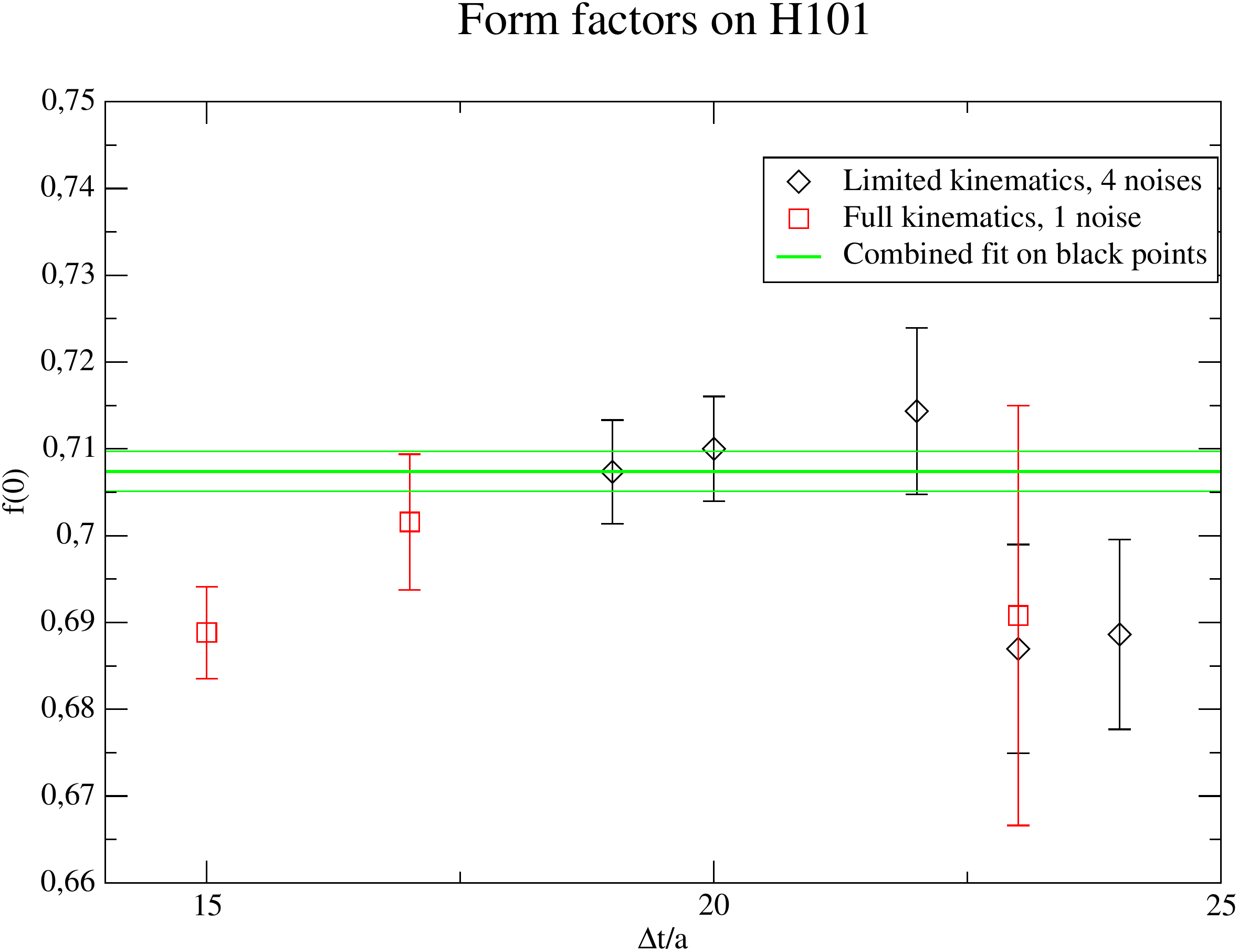}
  \end{center}
  \caption{\label{fig:f0deltaT}Points show the form factor computed from the double ratio. A small interpolation in $q^2$ is performed in each case.
  Those points are correlated since they share the same configurations and have sources and sinks relatively close. 
  Apart from the tension with the last two $\Delta t$, a clear tendency is visible which corresponds to the residual ${\cal O}(e^{-\Delta E \Delta t/2})$ contamination one would expect.
  The rightmost red point corresponds to Fig.~\ref{fig:H101deltaT}, while points with smaller source-sink separations are only used in this dedicated study and are cut from the rest of the analysis.}
\end{figure}

We therefore added a new set of runs dedicated to this question, with a minimal set of kinematics but a few $\Delta t$ and several $Z_2$ noises. We can determine the value of $f(0)$ computed with a single of those runs, and view it as a function of $\Delta t$ in Fig.~\ref{fig:f0deltaT}, or we can also compare with a combined fit of all 2-point and 3-point functions. 
This Bayes averaged combined fit is still a work in progress which suffers from some instabilities and poor $\chi^2$ values, but its preliminary results give interesting indications: it is compatible with most of the large-$\Delta t$ points and prefers single-exponential models, except if including the leftmost red points of Fig.~\ref{fig:f0deltaT} which do pick some two-exponential models. It also appears to have the potential to reduce the error bars.

\subsection{Taking the limits}

\begin{figure}[ht]
  \begin{center}
    \includegraphics[width=0.7\textwidth]{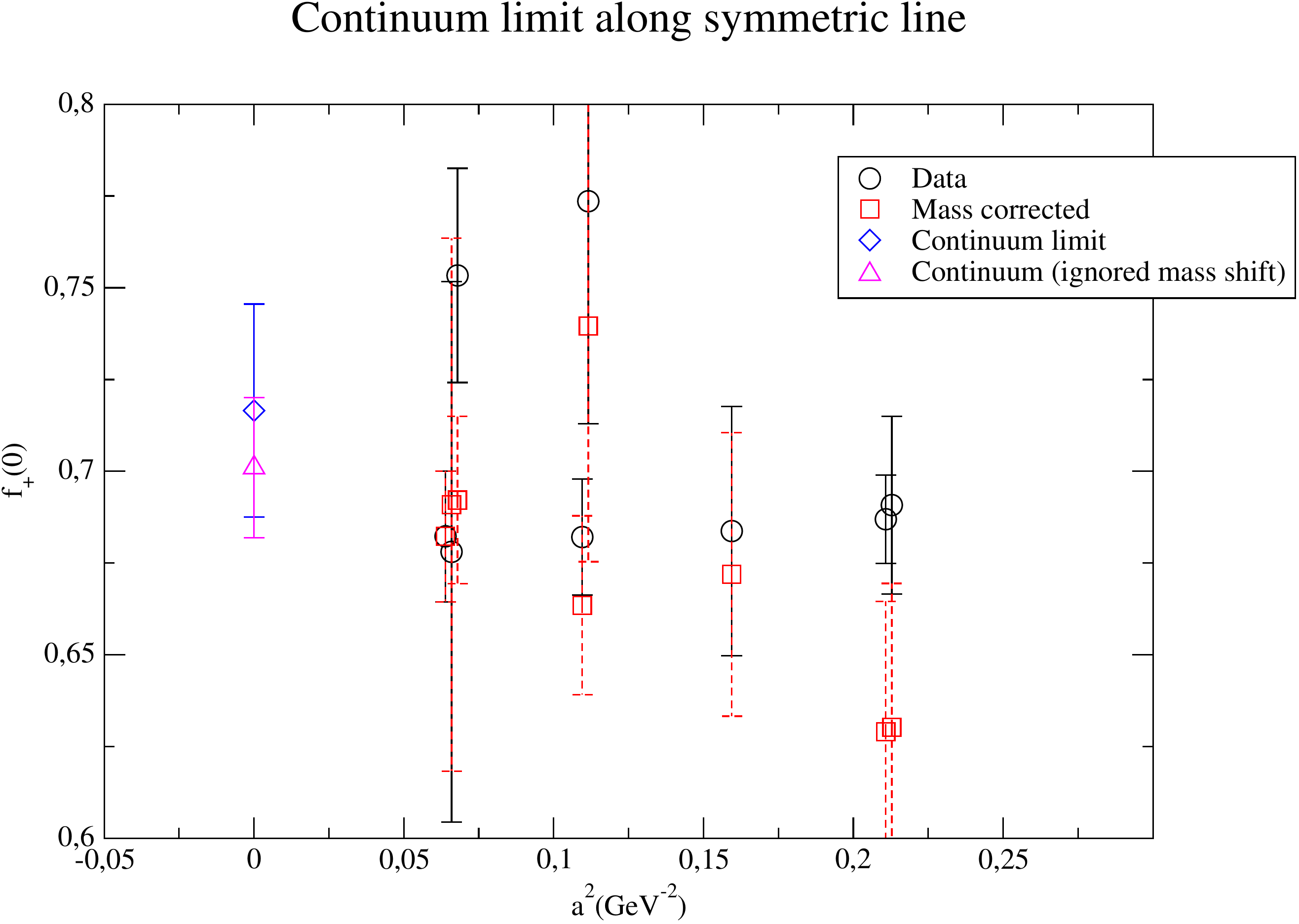}
  \end{center}
  \caption{\label{fig:f0-sym-cont}Subset of results for the particular case on $m_{ud}=m_s$ and $q^2=0$. In red the points are corrected for a slight mismatch of the charm mass, to keep $M_D/M_\pi$ constant.
  Continuum values are given from a correlated linear fit in $a^2$ (and $M_D/M_\pi$) with $\chi^2/dof = 0.37$(mass correction) or $1.14$ (correction ignored).}
\end{figure}

Now that we have looked in details at what happens for a specific ensemble, we can turn to the continuum limit. As shown in Fig.~\ref{fig:f0-sym-cont}, our current preliminary results look very compatible
with a linear fit in ${\cal O}(a^2)$ at this level of precision, with a $2\sigma$ signal on the slope. The value at the second coarsest ensemble is already compatible with the continuum limit, and the extrapolation only leads to a moderate increase of the error bars.

\begin{figure}[ht]
  \begin{center}
    \includegraphics[width=0.45\textwidth]{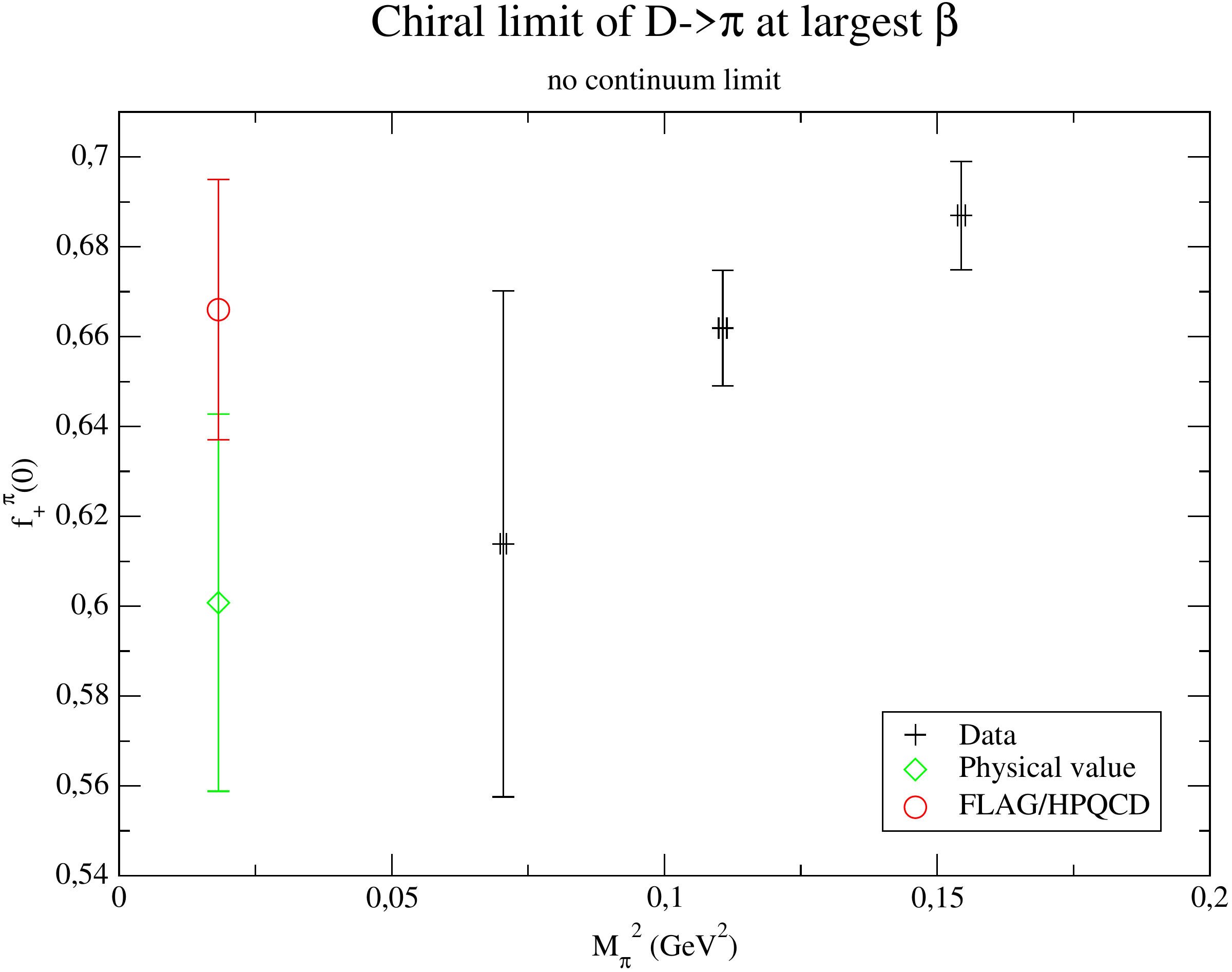}
    \hfill
    \includegraphics[width=0.45\textwidth]{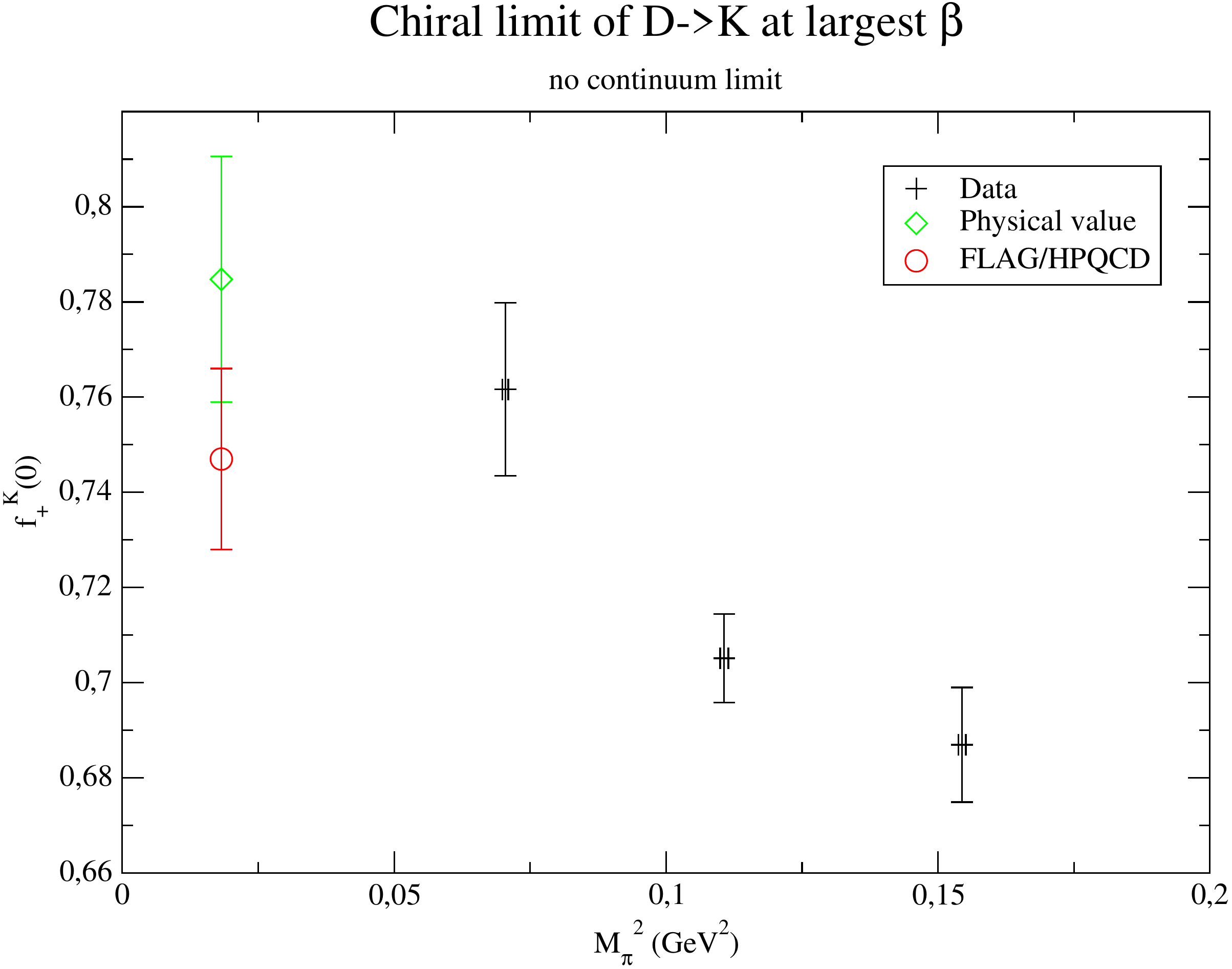}
  \end{center}
  \caption{\label{fig:f0chiral}Subset of results for the largest $\beta$ and $q^2=0$. A linear fit ($\chi^2/dof=0.16$ and $2.1$) is performed in $M_\pi^2$ and extrapolated to its physical value.
  The most recent published value from FLAG is given as a reference. This gives an indication on the potential of our method but one should not compare directly the central value since
  the continuum limit has been disregarded in this plot (see Fig.~\ref{fig:f0-sym-cont}). The leftmost point only uses 1 source noise, which we can easily improve in the future, while the other two have already been repeated with 4 noises.}
\end{figure}

Similarly, one can look at the mass dependence in Fig.~\ref{fig:f0chiral}. This time we can compare with FLAG results \cite{Aoki:2021kgd},
keeping in mind however that the continuum limit has not been taken into account.
For $D\to\pi$ we once again get a $2\sigma$ signal on the slope so that all point are almost compatible within error bars. We observe an increase of the error bars, and part of it could
be due to the fact that we need to inject larger momenta, but the main explanation is simply that this point has not yet accumulated the same number of $Z_2$ noises.
For $D\to K$, the mass dependence unexpectedly turns out to be more important: we obtain a $3\sigma$ for the slope and a mediocre $\chi^2/dof=2.1$. While this might only be a statistical fluctuation,
this calls for a comparison with other fitting models such as HMChPT, as well as some extra care with finite volume effects (which we can not distentangle given the precision we currently reach).

\section{Conclusion and perspectives}

We have presented preliminary results for charm semileptonic decays which turn out to give very sensible values, comparable in central value and error bar to what has already been published in the litterature.

The excited states appear to be fully under control once a conservative $2\ \fm$ cut is applied to the source-sink separation, but the cost to pay is a decrease in statistical precision. A more elaborate
combined fitting method is being developped to improve on that.
While, without this cut, excited state contaminations would break Lorentz symmetry, this can be disentangled from the ${\cal O}(a^2)$ effects observed by ETMc, which are insignificant with our action.

The discretisation effects are small even at zero squared momentum transfer, where large $3$-momenta are injected, and the continuum extrapolation seems to be under control. Nevertheless, future plans
include even finer ensembles with $a=0.039\ \fm$.

The mass dependence is relatively mild but might require additional work given the precision we are now reaching. Extra noise hits on H105 are already being computed. We also have results on finer low-$m_{ud}$ ensembles which we
chose not to present here because of their currently large error bars.

\acknowledgments
\noindent
We acknowledge PRACE and RES for giving us access to computational resources at MareNostrum (BSC). We thank CESGA for granting access to Finis Terrae II. More computing resources were also provided by
DESY, Zeuthen (PAX cluster).
This work is supported by the European Union's Horizon 2020 research and innovation programme under grant agreement No 813942 and by the Spanish MINECO through project PGC2018-094857-B-I00, the Centro de Excelencia Severo Ochoa Programme through SEV-2016-0597
and the Ramón y Cajal Programme RYC-2012-0249. We are grateful to CLS members for producing the gauge configuration ensembles used in this study.


\end{document}